\documentclass[twocolumn]{IEEEtran}

\usepackage{amsmath,bm}
\usepackage{amsthm}
\usepackage{amssymb}
\usepackage{graphicx}
\usepackage{epsfig}
\usepackage{psfrag}
\usepackage{subfigure}
\usepackage{url}
\usepackage{stfloats}
\usepackage{amsmath}
\usepackage{algorithm}
\usepackage{algorithmic}
\usepackage{diagbox}
\usepackage{dsfont}
\usepackage{bbding}

\usepackage{amsmath}
\usepackage{graphics}
\usepackage{graphicx}
\usepackage{amssymb}
\usepackage{algorithm}
\usepackage{algorithmic}
\usepackage{mathrsfs}
\usepackage{booktabs}
\usepackage{bm}
\usepackage{bbm}
\usepackage{float}
\usepackage{url}
\usepackage{amsfonts}
\usepackage[square,sort&compress,numbers]{natbib}
\usepackage{multirow}
\usepackage{textcomp}
\usepackage{supertabular}
\usepackage{array}
\usepackage{color,soul}

\begin{document}

\title{Pyramidal Dense Attention Networks for Lightweight Image Super-Resolution}

\author{ {Huapeng Wu, Jie Gui, \textit{Senior Member, IEEE}, Jun Zhang, \textit{Member, IEEE}, James T. Kwok, \textit{Fellow, IEEE}, and Zhihui Wei, \textit{Member, IEEE}}

\IEEEcompsocitemizethanks{
\IEEEcompsocthanksitem This work was supported in part by the National Natural
Science Foundation of China under Grant 11431015, Grant 61671243, Grant 61971223, Grant 61572463 and the Fundamental Research Funds for the Central Universities. Corresponding author: Zhihui Wei (e-mail: gswei@njust.edu.cn).
\IEEEcompsocthanksitem Huapeng Wu is with the School of Computer Science and Engineering, Nanjing University of Science and Technology, 210094, China.
\IEEEcompsocthanksitem Jie Gui is with the School of Cyber Science and Engineering, Southeast University, 211100, China.
\IEEEcompsocthanksitem Jun Zhang is with the School of Science, Nanjing University of Science and Technology, 210094, China.
\IEEEcompsocthanksitem James T. Kwok is with the Department of Computer Science and Engineering, Hong Kong University of Science and Technology, Hong Kong.
\IEEEcompsocthanksitem Zhihui Wei (corresponding author: gswei@njust.edu.cn) is with the School of Computer Science and Engineering, Nanjing University of Science and Technology, 210094, China, and with the School of Science, Nanjing University of Science and Technology, 210094, China.

} % \IEEEcompsocitemizethanks
} % \author

\markboth{}%
{Shell \MakeLowercase{\textit{et al.}}: Bare Demo of IEEEtran.cls for Journals}

\maketitle

\newcolumntype{L}[1]{>{\raggedright\arraybackslash}p{#1}}
\newcolumntype{C}[1]{>{\centering\arraybackslash}p{#1}}
\newcolumntype{R}[1]{>{\raggedleft\arraybackslash}p{#1}}

\begin{abstract}
	Recently, deep convolutional neural network methods have achieved an excellent performance in image super-resolution (SR), but they can not be easily applied to embedded devices due to large memory cost. To solve this problem, we propose a pyramidal dense attention network (PDAN) for lightweight image super-resolution in this paper. In our method, the proposed pyramidal dense learning can gradually increase the width of the densely connected layer inside a pyramidal dense block to extract deep features efficiently. Meanwhile, the adaptive group convolution that the number of groups grows linearly with dense convolutional layers is introduced to relieve the parameter explosion. Besides, we also present a novel joint attention to capture cross-dimension interaction between the spatial dimensions and channel dimension in an efficient way for providing rich discriminative feature representations. Extensive experimental results show that our method achieves superior performance in comparison with the state-of-the-art lightweight SR methods.
	
\end{abstract}


\section{Introduction}

Single image super-resolution (SISR) is a low-level vision problem that recovers a high resolution (HR) image from a low resolution (LR) observation, which is an ill-posed problem because multiple HR images can be degraded to the same LR image. To address this issue, researchers have proposed many approaches, which can be divided into three subclasses including interpolation-based methods \cite{zhou-et-al:scheme}, reconstruction-based methods \cite{tai-et-al:scheme}, and learning-based methods \cite{timofte2014a+, dong, 2016Deeply, 2017Image, 2017Deep, 2017Enhanced, 2020Wavelet, 2020Multi}.

Benefit from the powerful learning ability, various deep convolutional neural network based methods have been introduced and achieved a significant performance in image SR community \cite{2020Deep}. Firstly, Dong et al. \cite{dong} proposed a three-layer super-resolution convolutional neural network (SRCNN) to learn the nonlinear mapping function between LR and HR. Later, inspired by ResNet \cite{he2016deep} and DenseNet \cite{huang2017densely}, many complex deep neural networks \cite{zhang2018residual, dai2019second-order} have been presented to boost reconstruction performance. However, doing so will cause an increase in model parameters and computational cost, which greatly limit their practical applications in some computing devices, such as mobile and embedded applications. For solving these problems, some network architectures have been proposed by integrating recursive learning and lightweight models. For example, Deeply-Recursive Convolutional Network (DRCN) \cite{2016Deeply} and Deep Recursive Residual Network (DRRN) \cite{2017Image} used a recursive network to reduce the number of network parameters by parameter sharing strategies. Although DRCN and DRRN show favorable performances with a few parameters, they have a deeper model (DRRN ups to 52 convolutional layers) and require heavy computational costs. To further save computational overhead, some works proposed to directly extract feature in LR domain, and then upscale the features by deconvolution \cite{dong2016accelerating} or sub-pixel convolution \cite{shi2016real} at the end of the network. Afterward, Enhanced Deep Super-Resolution network (EDSR) \cite{2017Enhanced} adopted this post-processing strategy and empolyed a simplified ResNet architecture, which has about 165 convolutional layers. Zhang et al. \cite{zhang2018image} utilized residual-in-residual structure by adding channel attention block to construct a very deep model (up to 400 layers) for training a SR model. From these efforts, we find that the deeper network is helpful to extract more feature information to reconstruct HR images, but most of them suffered from large network parameters and computational cost.

Recently, some other researchers began to focus on design lightweight and efficient neural networks for image SR. Ahn et al. introduced a cascading residual network (CARN) \cite{2018Fast} to achieve efficient SR by using several cascading connections with group convolution \cite{2017Interleaved}. The information distillation network (IDN) \cite{hui} employed information distillation mechanism to divide the intermediate features into two parts, one part was retained and another part was further processed by succeeding convolution operations. To better balance performance and inference applications, Hui et al. \cite{2019Lightweight} further proposed a lightweight information multi-distillation network (IMDN) that extracted hierarchical features at a granular level, and split the preceding extracted features along channel dimension. At each step, its partial feature information was retained and the other features were processed in a subsequent step. It won the first place in the AIM 2019 constrained image super-resolution challenge \cite{AIM9022144}. In addition to the above-mentioned strategies, some other efficient methods (e.g. MobileNet \cite{2017MobileNets}, self-calibrated convolution \cite{2020Improving} and network architecture search \cite{2019Fast}) have also been proposed to design a lightweight deep neural network. Although these methods obtained comparable results, there is still room to obtain a better trade-off between model lightweight and SR performance.
\begin{figure*}
	\centering
	\includegraphics[width=\linewidth]{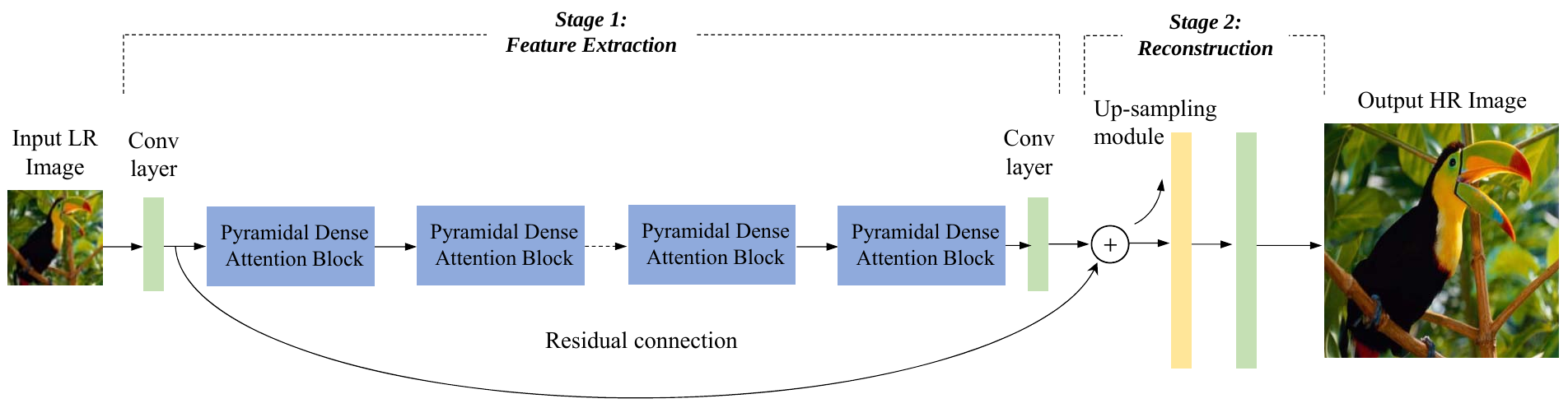}\\
	\caption{The architecture of our pyramidal dense attention network (PDAN). }\label{fig:PDAN}
\end{figure*}

In recent years, attention mechanisms have been widely used in many computer vision tasks and achieved good results \cite{zhang2018image, hu2019channel}. Hu et al. \cite{hu2018squeeze} proposed the squeeze-and-excitation block to learn the channel-wise information and improve the representation ability of the model. Inspired by \cite{hu2018squeeze}, Zhang et al. \cite{zhang2018image} introduced a residual channel attention network (RCAN) by integrating the squeeze-and-excitation block into the residual architecture for image SR. In \cite{2018CBAM}, the channel attention and spatial attention are combined to adaptively capture rich contextual information. Although these attention-based networks have provided performance improvements, they have some shortcomings. For example, SENet \cite{hu2018squeeze} only concentrate on channel-wise attention and ignore the importance of spatial information. Its variant, \cite{2018CBAM} uses the channel and spatial attention mechanisms, but they are only computed independently, which is also not beneficial to capture rich discriminative feature representations.

In this paper, we propose a novel lightweight pyramidal dense attention network (PDAN) for SISR (illustrated in Fig. \ref{fig:PDAN}). Unlike most previous dense networks with a fixed channel growth rate, we introduce a lightweight pyramidal dense block with a various channel growth rate inspired by \cite{2020MC}. The proposed feature extractor adopts the pyramidal dense learning strategy that the output channel dimensionality grows up while the convolutional layers are deepening in each pyramidal dense block, which can effectively aggregate contextual information (shown in Fig. \ref{fig:PDAB}). While increasing the output feature dimensionality is beneficial to enhance the feature learning ability of the networks, the parameter explosion needs to be noted. Therefore, we use group convolution with the dimension cardinality (i.e. the number of groups in each convolutional layer) which increases linearly to relieve the parameter explosion. Besides, to further improve the discriminative representation ability of the network, we propose a novel joint attention to capture cross-dimension interaction between the channel dimension and the spatial dimensions \cite{Misra_2021_WACV} for high-frequency information extraction. The detail of joint attention is shown in Fig. \ref{fig:PDAB} (b), where an input tensor (the size is $C \times H \times W $) is transferred to four branches, two of which are responsible for modeling channel attention ($C \times 1 \times 1$) and spatial attention ($H, W$) respectively, and the other two branches are used to build cross-dimension interaction between the channel dimension $C$ and the spatial dimension $H$ or $W$.

The main contributions of this paper are summarized as follows:

(1)	We propose a lightweight pyramidal dense attention network (PDAN) for SISR. The proposed lightweight model can effectively improve SR performance via the pyramidal dense feature learning and joint attention mechanism with a few parameters.

(2)	We introduce a pyramidal dense block to fully exploit multi-level features as well as boost the flow of information with a few parameters. Moreover, the proposed joint attention can efficiently capture cross-dimension interaction between the channel dimension and the spatial dimensions to provide richer and more discriminative contextual information for feature rescaling. In comparison with state-of-the-art lightweight SR methods, our method achieves competitive results in terms of model size and performance.

The remainder of this paper is organized as follows. In Section II, we introduce the proposed PDAN for image SR in detail. Extensive experimental results and analysis are provided in Section III. Finally, the conclusion is given in Section IV.

\section{Proposed Method} \label{sec:Methodology}

In this section, we will show the proposed lightweight pyramidal dense attention networks for SISR in details.

\subsection{Network Architecture}

As illustrated in Fig. \ref{fig:PDAN}, the proposed PDAN consists of two modules: 1) feature extraction module, 2) reconstruction module. In this paper, we denote $I^{LR}$ and $I^{SR}$ as the input LR image and output HR image of our PDAN. The LR image is first fed to one ${\rm{3}} \times {\rm{3}}$ convolution layer for initial feature extraction. Next, the learned initial features are fed into several stacked lightweight pyramidal dense attention blocks to produce powerful feature maps. Lastly, we apply the reconstruction module that consists of a ${\rm{3}} \times {\rm{3}}$  convolution layer and a sub-pixel convolution layer \cite{shi2016real} to reconstruct the desired SR image. The SR procedure of the proposed PDAN can be written as following:
\begin{equation}
{I^{SR}} = {F_{PDAN}}\left( {{I^{LR}, \theta}} \right) = {F_{UP}}\left({F_{e}}\left({{I^{LR}}} \right) \right),
\end{equation}
where ${F_{PDAN}}$  represents the reconstruction function of our PDAN, $\theta$  denotes the learnable parameters of the network, ${F_{e}}$  is the feature extraction step and  $F_{up}$ denotes the up-sampling and reconstruction step.

Our PDAN is optimized with the  $L_1$ loss \cite{zhang2018image} by measuring the difference between a reconstructed SR image $I^{SR}$  and its HR ground-truth $I^{HR}$ . Given a set of training image pairs $\left\{ {I_i^{LR},I_i^{HR}} \right\}_{i = 1}^N$ , where $N$ denotes the number of training images. The loss function of our method can be expressed as
\begin{equation}
L(\theta) = \frac{1}{N}\sum_{i=1}^N \| F_{PDAN}(I^{LR}_i, \theta) - I^{HR}_i\|_1.
\end{equation}
Next, we will give more details about the proposed lightweight pyramidal dense attention block.

\begin{figure*}
	\centering
	\includegraphics[height= 6in, width=\linewidth]{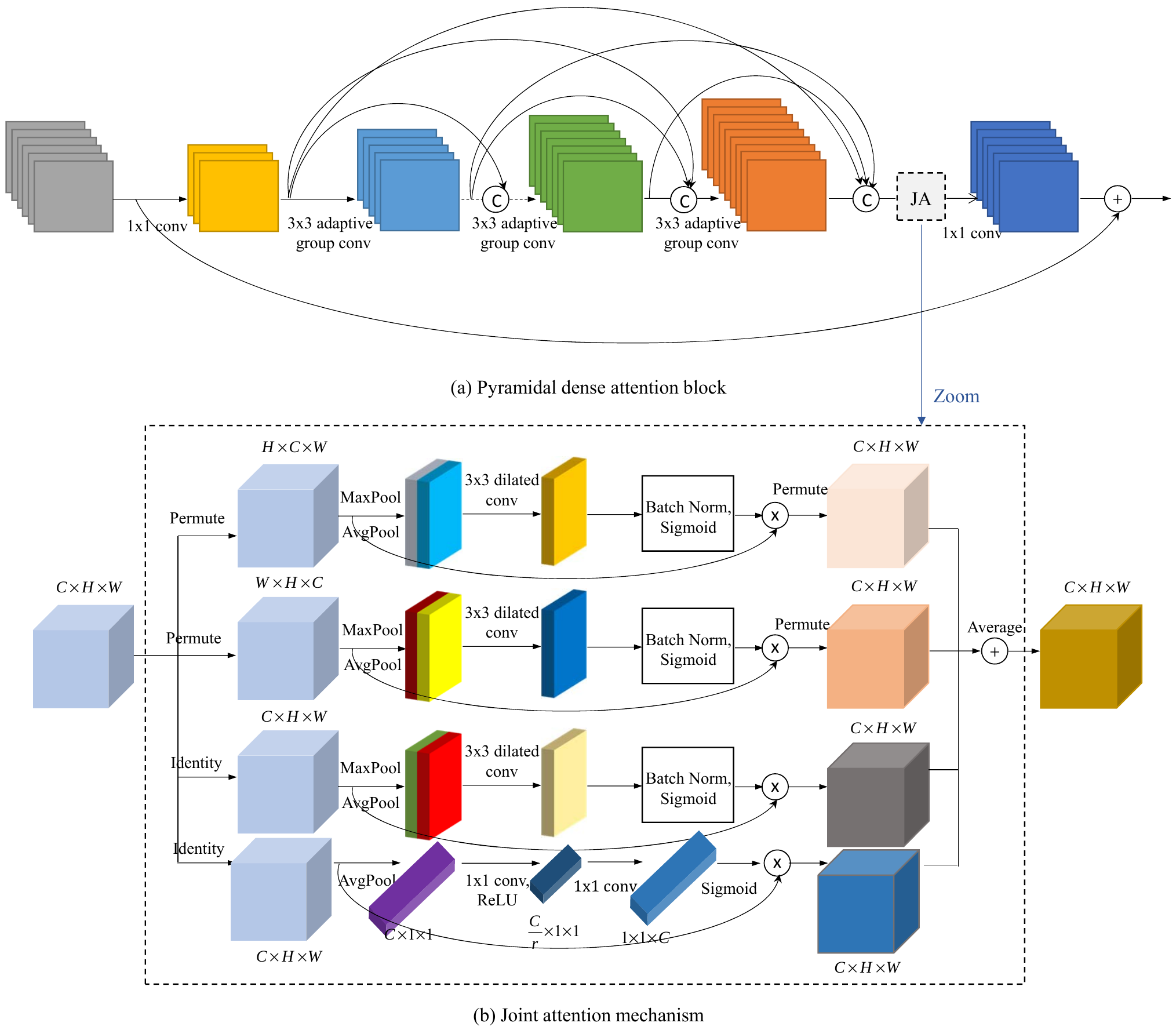}\\
	\caption{(a) The illustration of the proposed pyramidal dense attention block. (b) The details of the joint attention module.}\label{fig:PDAB}
\end{figure*}

\subsection{Pyramidal Dense Attention Block (PDAB)}

As illustrated in Fig. \ref{fig:PDAB} (a), our pyramidal dense attention block (PDAB) is mainly constructed by two stages: pyramidal dense feature learning, and feature refinement based on joint attention mechanism. Moreover, local residual learning is also adopted to improve the gradient information flow. Let  $H_{d-1}$ and  ${H_d}$ denote the input and output of the $d$-th PDAB, respectively. The process flow with PDAB can be formulated as following:
\begin{equation}
H_d = H_{d-1} + F_{PDAB}(H_{d-1}),
\end{equation}
where  $F_{PDAB}$ is the mapping function of the proposed PDAB.

\subsubsection{Pyramidal Dense Feature Learning}

Dense skip connections have been widely exploited in deep learning methods, which is beneficial to boost the flow of feature information and enhance the representation ability of the network. However, on the other hand, the dense connections will need large resources and memory for extracting and storing features. Additionally, due to the inconsistency between the number of input and output channels, the dense connections with a fixed channel growth rate can cause the loss of information and degrade the learning ability of the network. In this paper, we introduce a pyramidal dense connection mechanism with a varying growth rate, where the output feature dimensionality of convolutional layers gradually increases when the network layers are deepening to relieve this problem. However, when the convolution layers increase, this method will result in the parameter explosion. Specifically, the input channel dimensionality of two different dense connection layers can be written as follows.

\noindent The conventional dense connections with a fixed growth rate are
\begin{equation}
C_i = c_0 + \sum_{j=1}^{i-1} c_j = c_0 + (i-1) \times g_0 ,
\end{equation}
\begin{equation}
c_j = g_0.
\end{equation}

\noindent The pyramidal dense connections with a varying growth rate are
\begin{equation}
\widetilde C_i = c_0 + \sum_{j=1}^{i-1} \widetilde c_j = c_0 + (i-1) \times g_0 + g \times \frac{(i-1)(i-2)}{2} ,
\end{equation}
\begin{equation}
\widetilde c_j = g_0 + g \times (j-1),
\end{equation}
where $c_0$ is the number of initial input channels, $g$ is a increasing factor of growth rate $g_0$. $C_i$ and $\widetilde C_i$ are the number of input channels in the $i$-th layer, $c_j$ and $\widetilde c_j$ denote the output feature dimensionality of $j$-th convolutional layer.

From the above analysis, the group convolution is adopted to avoid the problem of parameter explosion. Note that we choose group convolution instead of conventional convolution since the former can reduce the number of parameters efficiently. Specifically, the corresponding parameters of them are given as
\begin{equation}
P_i = C^{out}_i \times C^{in}_i \times K \times K,
\end{equation}
\begin{equation}
P^G_i = G_i \times (\frac{C^{out}_i}{G_i} \times \frac{C^{in}_i}{G_i} \times K \times K) = \frac{P_i}{G_i},
\end{equation}
where $C^{in}_i$ and $C^{out}_i$ are the number of input and output channels in the $i$-th layer, $K$ indicates the corresponding kernel size, $G_i$ is the group size and the bias is omitted for simplicity. Obviously, the parameters of group convolution ($P^G$) are less than the parameters of conventional convolution ($P$). However, when $G_i$ is fixed, this will also cause the parameter to increase sharply since the output feature channels $C^{out}$ grows up linearly with the dense layers. Therefore, we introduce the adaptive group convolution to further alleviate the parameter explosion. In the experiment, the group size is set to grow linearly with dense convolutional layers,
\begin{equation}
G_i = i+1.
\end{equation}

Moreover, the bottleneck architecture ($1 \times 1$ convolutional filters) is also designed to refine the input cascaded features to be consistent with the output channels of the corresponding layer, which can keep them divisible by the group size and boost the information propagation among different group features in each layer.

\subsubsection{Joint Attention Mechanism (JA)}

The aim of the attention-based SR methods is to selectively focus on more important high-frequency details of the input feature for HR image recovery. Here, we introduce a novel joint attention that can capture cross-dimension dependencies \cite{Misra_2021_WACV} for improving the discriminative feature representations. First, we will revisit some attention mechanisms including channel attention and spatial attention \cite{hu2018squeeze,2018CBAM}.

\textbf{Convolutional Block Attention Module}: Unlike SENet \cite{hu2018squeeze}, convolutional block attention module (CBAM) \cite{2018CBAM} compute channel attention by using global average pooling and global max pooling. Specifically, let $X\in \mathbb{R}^{C\times H \times W}$ denote the input feature where $C$ is the number of input channels, $H$ and $W$ are height and width of the input features, respectively. The channel-wise global statistics are defined as
\begin{equation}
GAP(X) = \frac{1}{H\times W}\sum_{i=1}^H\sum_{j=1}^W X_{i,j},
\end{equation}
\begin{equation}
GMP(X) = \max\limits_{H,W}\ (X_{i,j}).
\end{equation}
Next, the multi-layer perception is shared to compute channel attention,
\begin{equation}
A_e = \sigma(W_2\delta(W_1GAP(X))+W_2\delta(W_1GMP(X))),
\end{equation}
where $\sigma$ and $\delta$ are the sigmoid function and the Rectified Linear Unit \cite{2011Deep}, $W_1\in \mathbb{R}^{\frac{C}{r}\times C \times 1 \times 1}$, $W_2\in \mathbb{R}^{C \times \frac{C}{r} \times 1 \times 1}$, and $r$ is the reduction radio. Moreover, the above average pooling and max pooling operations along the channel dimension are also used to produce the spatial attention weights in CBAM. The formulation can be written by
\begin{equation}
A_s = \sigma(w^{k \times k}([X^{avg}_s, X^{max}_s])),
\end{equation}
where $X^{avg}_s$, $X^{max}_s\in \mathbb{R}^{1 \times H \times W}$ , $W^{k \times k}$ denotes a convolution layer with the filter size of $k \times k$  and $\sigma$ is the sigmoid function.

\textbf{Joint Attention module}: Although CBAM \cite{2018CBAM} introduces channel attention and spatial attention to improve the feature representations, they ignore the importance of cross-dimension interaction. For this problem, we propose the joint attention to effectively model channel attention and spatial attention with few parameters.

As shown in Fig. \ref{fig:PDAB} (b), joint attention consists of four branches, two branches are responsible for modeling the channel attention ($C \times 1 \times 1$) and the spatial attention ($H, W$). The rest two branches aim to capture the interaction between the channel dimension and the spatial dimensions ($(H, C)$ or $(C,W)$). Specifically, given a input feature map $X \in \mathbb{R}^{C \times H \times W}$. In the channel attention branch, like SENet \cite{hu2018squeeze}, the channel attention weights are used to refine the feature $\hat X_{C}$. In the spatial attention branch, we first aggregate channel information to produce two 2D maps ($1 \times H \times W$ ) by the above average-pooling and max-pooling operations across the channel axis. Then, a $3 \times 3$ dilated convolution \cite{2016Multi} with dilation value 3 followed by a batch normalization \cite{2015Batch} and a sigmoid activation function are used for producing the attention weights ($1 \times H \times W$). Finally, the spatial attention weights are multiplied with the input tensor $X$ to generate the recalibrated feature $\hat X_{H \times W}$. For the remaining two branches, the input tensor $X$ is first permuted and then passed to the above similar operations to generate the corresponding attention weights ($1 \times H \times C$ or $1 \times C \times W$  ), respectively. Subsequently, the generated attention weights are applied on the permuted feature and then rotated ${90^ \circ }$ clockwise along the width axis or the height axis to retain the original input shape ($C \times H \times W$). Finally, the refined features of four branches ($\hat X_{C}$, $\hat X_{H \times W}$, $\hat X_{H \times C}$, $\hat X_{C \times W}$) are aggregated by the averaging operation to enhance the discriminative representation ability of the network, which can be written as
\begin{equation}
\hat X = \frac {\hat X_{C} + \hat X_{H \times W} + \hat X_{H \times C} + \hat X_{C \times W}} {4},
\end{equation}
where $\hat X$ is the output recalibrated feature. In addition to simple averaging, other simple and efficient aggregation methods can also be explored to further improve the performance.

\section{Experimental Results and Analysis} \label{sec:Experiments}

\subsection{Datasets and Metrics}

In our experiments, we use DIV2K dataset \cite{timofte2017ntire} as our training set, which contains 800 high-quality training images. We employ five standard benchmark datasets for testing our model, including Set5 \cite{bevilacqua2012low}, Set14 \cite{2010On}, Bsd100 \cite{2002A}, Urban100 \cite{2015Single} and Manga109 \cite{2017Sketch}. We conduct three different experiments with bicubic (BI),
blur-downscale (BD) and downscale-noise (DN) degradation models. The peak signal-to-noise ratio (PSNR) and the structural similarity index measurement (SSIM) \cite{wang2004image} are used for measuring the quality of the SR images on the luminance (Y) channel of the transformed YCbCr space.

\begin{table*}
	\newcommand{\tabincell}[2]{\begin{tabular}{@{}#1@{}}#2\end{tabular}}
	\centering
	\caption{Ablation study of different combinations in the method. We report the parameters, FLOPs and PSNR on Manga109 ($\times {\rm{4}}$).}\label{ablation}
	\begin{tabular}{l|C{1.2cm}|C{1.2cm}|C{1.2cm}|C{1.2cm}}
		\toprule
		& ${P_{\rm{1}}}$ & ${P_{\rm{2}}}$ & ${P_{\rm{3}}}$ & ${P_{\rm{4}}}$ \\
		\midrule
		Channel attention \cite{hu2018squeeze} & & \checkmark & & \\
		\hline
		Channel and Spatial attention \cite{2018CBAM} & & & \checkmark & \\
		\hline
		Joint attention & & & & \checkmark \\
		\hline
		Params (K) & 1471 & 1586 & 1588 & 1587 \\
		\hline
		FLOPs (G) & 31.78 & 31.85 & 31.87 & 31.87 \\
		\hline
		PSNR (dB) & 30.47 & 30.59 & 30.46 & 30.64 \\
		\bottomrule
	\end{tabular}
\end{table*}

\subsection{Implementation Details}

To construct the training pairs, we downsample the original HR images to generate LR Images by using bicubic interpolation. Data augmentation is performed on the above training dataset by randomly rotating ${90^ \circ }$, ${\rm{18}}{{\rm{0}}^ \circ}$, ${\rm{27}}{{\rm{0}}^ \circ }$ and flipping horizontally. In each training batch, 16 LR RGB patches with the size of $48 \times 48$ are randomly sampled as inputs. We adopt Adam optimizer \cite{2014Adam} to train the model by setting ${\beta _{\rm{1}}}{\rm{ = 0}}{\rm{.9}}$, ${\beta _{\rm{2}}}{\rm{ = 0}}{\rm{.999}}$, and $\varepsilon {\rm{ = 1}}{{\rm{0}}^{ - 8}}$. The learning rate is initialized as ${10^{ - 4}}$ and reduced to half every 200 epochs. We implement the proposed method by using the PyTorch framework \cite{paszke2017automatic} with a NVIDIA RTX 2080Ti GPU.

In our proposed model, the convolutional layers are set to 64 filters with $3 \times 3$ kernel size except for $1 \times 1$ convolutional layers for feature fusion. Additionally, in each pyramidal dense block, the initial input features are first reduced to $c_0$ and then passed to the subsequent pyramidal dense connection layers. For simplicity, the channel number of pyramidal dense layers can be expressed linearly as $c_j = (j+1) \times c_0$.  The operations of each pyramidal dense layer consists of a $1 \times 1$ convolution filter, activation function ReLU, $3 \times 3$ adaptive group convolution and ReLU. Specifically, $c_0$ is set to 16 and the number of pyramidal dense layers is 4 in a pyramidal dense block. Following the previous papers, we use the recently popular sampling method ESPCNN \cite{shi2016real} to perform the upsampling operation. The final convolution layer has 3 filters with kernel size of $3 \times 3$ that are used to reconstruct the HR image.

\begin{figure*}
	\centering
	\includegraphics[width=\linewidth]{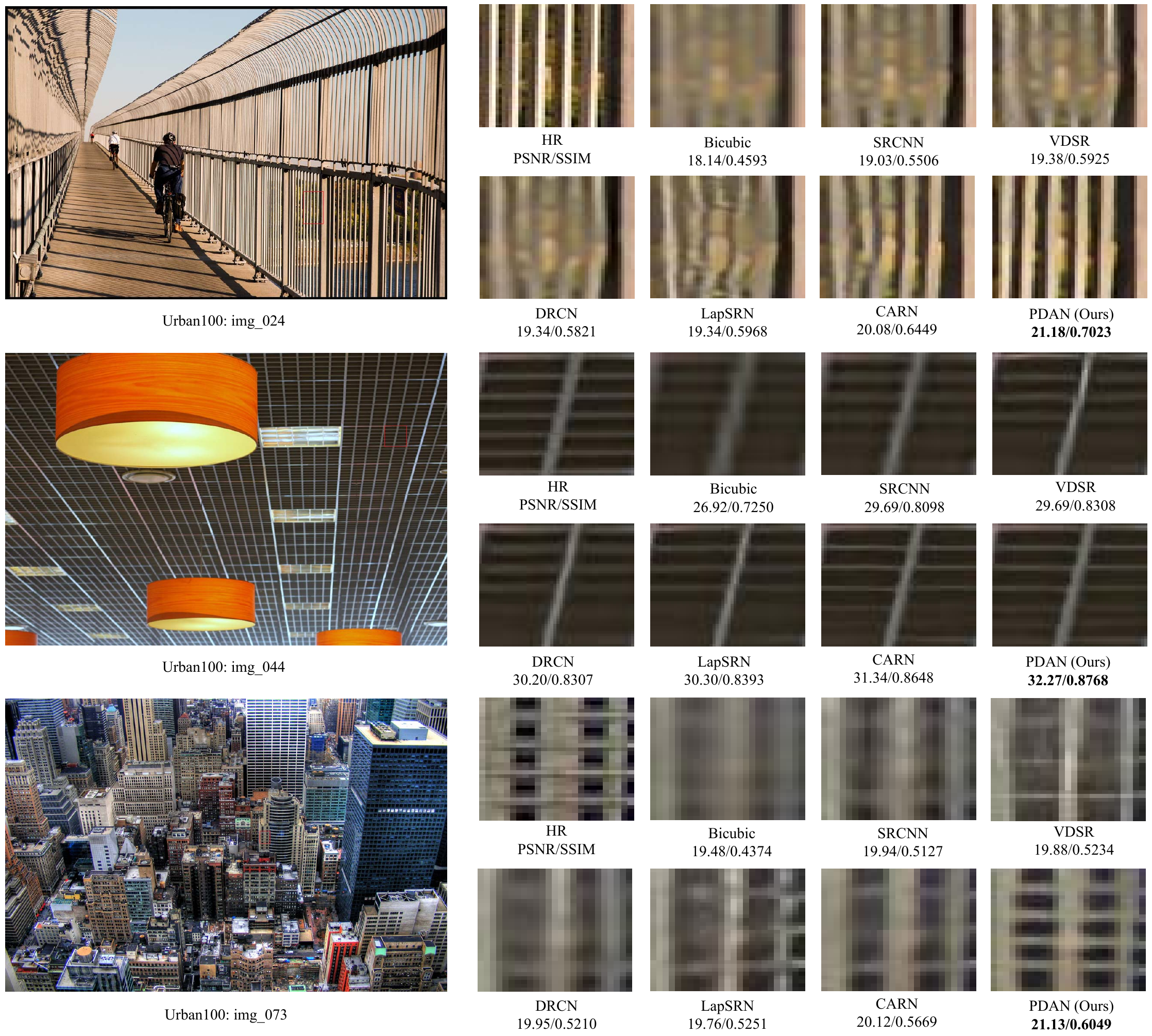}\\
	\caption{Visual comparison for $\times 4$ SR with BI model on Urban100 dataset. The best results are highlighted.
	}\label{fig:urban}
\end{figure*}

\subsection{Ablation Study}

In this subsection, we analyze the effects of the key configurations in our proposed method. For fair comparisons, our method and its variants are set with the same baseline structure. As shown in Table \ref{ablation}, we can see that the baseline method $P_1$ (without any attention) performs not very well and PSNR only reaches 30.47 dB on Manga109 ($\times 4$). It is noticed that when we compare the joint attention ($P_4$) with other attention mechanisms ($P_2$ and $P_3$), we can see that our joint attention module can achieve better performance (i.e., 30.64 v.s. 30.59 dB v.s. 30.46 dB).

In addition, we also compare other metrics, including parameters and FLOPs to demonstrate the effectiveness of our method. FLOPs is computed on the HR image with $512 \times 512$ pixels. From Table \ref{ablation}, it is observed that our joint attention model can improve the PSNR with less additional parameters and FLOPs. The results clearly indicate that joint attention is more effective than other attention mechanisms \cite{hu2018squeeze,2018CBAM}.

\begin{table*}
	\newcommand{\tabincell}[2]{\begin{tabular}{@{}#1@{}}#2\end{tabular}}
	\centering
	\caption{Quantitative results of different image SR methods with BI degradation module.The parameters and PSNR/SSIM comparison results for scaling factor $ \times {\rm{2}}$, $ \times {\rm{3}}$ and $ \times {\rm{4}}$. Red/blue text: best/second-best among all methods.}\label{comp_with_others_BI}
	\begin{tabular}{L{1.8cm}|C{0.8cm}|C{0.8cm}|C{0.8cm}C{0.8cm}|C{0.8cm}C{0.8cm}|C{0.8cm}C{0.8cm}|C{0.8cm}C{0.8cm}|C{0.8cm}C{0.8cm}}
		\toprule
		\multicolumn{1}{c}{} & \multicolumn{1}{c}{} & \multicolumn{1}{c}{} &
		\multicolumn{2}{c}{Set5} & \multicolumn{2}{c}{Set14} & \multicolumn{2}{c}{Bsd100} & \multicolumn{2}{c}{Urban100} & \multicolumn{2}{c}{Manga109}\\
		Methods & Scale & Params & PSNR & SSIM & PSNR & SSIM & PSNR & SSIM & PSNR & SSIM & PSNR & SSIM \\
		\midrule
		Bicubic & 2 & $-$ & 33.66 & 0.9299 & 30.24 & 0.8688 & 29.56 & 0.8431 & 26.88 & 0.8403 & 30.80 & 0.9339 \\
		SRCNN \cite{dong} & 2 & 57K & 36.66 & 0.9542 & 32.45 & 0.9067 & 31.36 & 0.8879 & 29.50 & 0.8946 & 35.60 & 0.9663 \\
		FSRCNN \cite{dong2016accelerating} & 2 & 13K & 37.00 & 0.9558 & 32.63 & 0.9088 & 31.53 & 0.8920 & 29.88 & 0.9020 & 36.67 & 0.9710 \\
		VDSR \cite{kim2016accurate} & 2 & 666K & 37.53 & 0.9587 & 33.03 & 0.9124 & 31.90 & 0.8960 & 30.76 & 0.9140 & 37.22 & 0.9750 \\
		DRCN \cite{2016Deeply} & 2 & 1774K & 37.63 & 0.9588 & 33.04 & 0.9118 & 31.85 & 0.8942 & 30.75 & 0.9133 & 37.55 & 0.9732 \\
		LapSRN \cite{lai2017deep} & 2 & 251K & 37.52 & 0.9591 & 32.99 & 0.9124 & 31.80 & 0.8952 & 30.41 & 0.9103 & 37.27 & 0.9740 \\
		DRRN \cite{2017Image} & 2 & 298K & 37.74 & 0.9591 & 33.23 & 0.9136 & 32.05 & 0.8973 & 31.23 & 0.9188 & 37.88 & 0.9749 \\
		SRMDNF \cite{zhang2018learning} & 2 & 1511K & 37.79 & 0.9601 & 33.32 & 0.9159 & 32.05 & 0.8985 & 31.33 & 0.9204 & 38.07 & 0.9761 \\
		SRResNet \cite{ledig2017photo} & 2 & 1370K & \textcolor{blue}{38.05} & \textcolor{blue}{0.9607} & \textcolor{blue}{33.64} & \textcolor{blue}{0.9178} & \textcolor{red}{32.22} & \textcolor{red}{0.9002} & \textcolor{blue}{32.23} & \textcolor{blue}{0.9295} & 38.05 & 0.9607 \\
		CARN \cite{2018Fast} & 2 & 1592K & 37.76 & 0.9590 & 33.52 & 0.9166 & 32.09 & 0.8978 & 31.92 & 0.9256 & \textcolor{blue}{38.36} & \textcolor{blue}{0.9765} \\
		PDAN (ours) & 2 & 1439K & \textcolor{red}{38.05} & \textcolor{red}{0.9607} & \textcolor{red}{33.65} & \textcolor{red}{0.9182} & \textcolor{blue}{32.20} & \textcolor{blue}{0.8998} & \textcolor{red}{32.36} & \textcolor{red}{0.9300} & \textcolor{red}{38.71} & \textcolor{red}{0.9771} \\
		\midrule
		Bicubic & 3 & $-$ & 30.39 & 0.8682 & 27.55 & 0.7742 & 27.21 & 0.7385 & 24.46 & 0.7349 & 26.95 & 0.8556 \\
		SRCNN \cite{dong} & 3 & 57K & 32.75 & 0.9090 & 29.30 & 0.8215 & 28.41 & 0.7863 & 26.24 & 0.7989 & 30.48 & 0.9117 \\
		FSRCNN \cite{dong2016accelerating} & 3 & 13K & 33.18 & 0.9140 & 29.37 & 0.8240 & 28.53 & 0.7910 & 26.43 & 0.8080 & 31.10 & 0.9210 \\
		VDSR \cite{kim2016accurate} & 3 & 666K & 33.66 & 0.9213 & 29.77 & 0.8314 & 28.82 & 0.7976 & 27.14 & 0.8279 & 32.01 & 0.9340 \\
		DRCN \cite{2016Deeply} & 3 & 1774K & 33.82 & 0.9226 & 29.76 & 0.8311 & 28.80 & 0.7963 & 27.15 & 0.8276 & 32.24 & 0.9343 \\
		LapSRN \cite{lai2017deep} & 3 & 502K & 33.81 & 0.9220 & 29.79 & 0.8325 & 28.82 & 0.7980 & 27.07 & 0.8275 & 32.21 & 0.9350 \\
		DRRN \cite{2017Image} & 3 & 298K & 34.03 & 0.9244 & 29.96 & 0.8349 & 28.95 & 0.8004 & 27.53 & 0.8378 & 32.71 & 0.9379 \\
		SRMDNF \cite{zhang2018learning} & 3 & 1528K & 34.12 & 0.9254 & 30.04 & 0.8382 & 28.97 & 0.8025 & 27.57 & 0.8398 & 33.00 & 0.9403 \\
		SRResNet \cite{ledig2017photo} & 3 & 1554K & \textcolor{blue}{34.41} & \textcolor{blue}{0.9274} & \textcolor{blue}{30.36} & \textcolor{blue}{0.8427} & \textcolor{blue}{29.11} & \textcolor{blue}{0.8055} & \textcolor{blue}{28.20} & \textcolor{blue}{0.8535} & \textcolor{blue}{33.54} & \textcolor{blue}{0.9448} \\
		CARN \cite{2018Fast} & 3 & 1592K & 34.29 & 0.9255 & 30.29 & 0.8407 & 29.06 & 0.8034 & 28.06 & 0.8493 & 33.50 & 0.9440 \\
		PDAN (ours) & 3 & 1624K & \textcolor{red}{34.44} & \textcolor{red}{0.9276} & \textcolor{red}{30.39} & \textcolor{red}{0.8437} & \textcolor{red}{29.11} & \textcolor{red}{0.8063} & \textcolor{red}{28.34} & \textcolor{red}{0.8563} & \textcolor{red}{33.63} & \textcolor{red}{0.9448} \\
		\midrule
		Bicubic & 4 & $-$ & 28.42 & 0.8104 & 26.00 & 0.7027 & 25.96 & 0.6675 & 23.14 & 0.6577 & 24.89 & 0.7866 \\
		SRCNN \cite{dong} & 4 & 57K & 30.48 & 0.8628 & 27.50 & 0.7513 & 26.90 & 0.7101 & 24.52 & 0.7221 & 27.58 & 0.8555 \\
		FSRCNN \cite{dong2016accelerating} & 4 & 12K & 30.72 & 0.8660 & 27.61 & 0.7550 & 26.98 & 0.7150 & 24.62 & 0.7280 & 27.90 & 0.8610 \\
		VDSR \cite{kim2016accurate} & 4 & 665K & 31.35 & 0.8830 & 28.02 & 0.7680 & 27.29 & 0.7251 & 25.18 & 0.7540 & 28.83 & 0.8870 \\
		DRCN \cite{2016Deeply} & 4 & 1774K & 31.53 & 0.8854 & 28.02 & 0.7670 & 27.23 & 0.7233 & 25.14 & 0.7510 & 28.98 & 0.8816 \\
		LapSRN \cite{lai2017deep} & 4 & 813K & 31.54 & 0.8850 & 28.19 & 0.7720 & 27.32 & 0.7270 & 25.21 & 0.7560 & 29.09 & 0.8900 \\
		DRRN \cite{2017Image} & 4 & 297K & 31.68 & 0.8888 & 28.21 & 0.7720 & 27.38 & 0.7284 & 25.44 & 0.7638 & 29.45 & 0.8946 \\
		SRMDNF \cite{zhang2018learning} & 4 & 1552K & 31.96 & 0.8925 & 28.35 & 0.7787 & 27.49 & 0.7337 & 25.68 & 0.7731 & 30.09 & 0.9024 \\
		SRResNet \cite{ledig2017photo} & 4 & 1518K & \textcolor{blue}{32.17} & \textcolor{blue}{0.8951} & \textcolor{blue}{28.61} & \textcolor{blue}{0.7823} & \textcolor{blue}{27.59} & \textcolor{blue}{0.7365} & \textcolor{blue}{26.12} & \textcolor{blue}{0.7871} & \textcolor{blue}{30.48} & \textcolor{blue}{0.9087} \\
		CARN \cite{2018Fast} & 4 & 1592K & 32.13 & 0.8937 & 28.60 & 0.7806 & 27.58 & 0.7349 & 26.07 & 0.7837 & 30.47 & 0.9084 \\
		PDAN (ours) & 4 & 1587K & \textcolor{red}{32.28} & \textcolor{red}{0.8957} & \textcolor{red}{28.66} & \textcolor{red}{0.7831} & \textcolor{red}{27.62} & \textcolor{red}{0.7378} & \textcolor{red}{26.27} & \textcolor{red}{0.7922} & \textcolor{red}{30.64} & \textcolor{red}{0.9098} \\
		\bottomrule
	\end{tabular}
\end{table*}

\subsection{Results with Bicubic Degradation (BI)}

We compare our proposed method with some state-of-the art SR methods, Bicubic, SRCNN \cite{dong}, FSRCNN \cite{dong2016accelerating}, VDSR \cite{kim2016accurate}, DRCN \cite{2016Deeply}, LapSRN \cite{lai2017deep}, DRRN \cite{2017Image}, SRMDNF \cite{zhang2018learning}, SRResNet \cite{ledig2017photo} and CARN \cite{2018Fast}. Table \ref{comp_with_others_BI} shows quantitative results for three different scaleing factors ($\times 2$, $\times 3$, $\times 4$). In addition, the number of parameters of these methods is also provided to show the model complexity. The results show that our PDAN achieves the comparable results on all the datasets and scaling factors in terms of both PSNR and SSIM. Although CARN \cite{2018Fast} has similar parameters as us, our PDAN can obtains superior performance. For example, when the scaling factor is $\times 4$, the average PSNR gain of our PDAN over the CARN \cite{2018Fast} are 0.20 dB, 0.17 dB on Urban100 and Manga109 datasets, respectively.

In Fig. \ref{fig:urban}, we present visual comparison with upscaling factor $\times 4$ on Urban100 dataset. We can see that that our method yields best visual results among all existing compared methods. For ``$image\_024$'', the other compared methods generate the blurred detailed textures. Our method can recover more accurate lines and details. For ``$image\_073$'', most methods suffer from the blurry effects while our method produces the sharper edges and finer details.

\begin{figure}
	\centering
	\includegraphics[width=\linewidth]{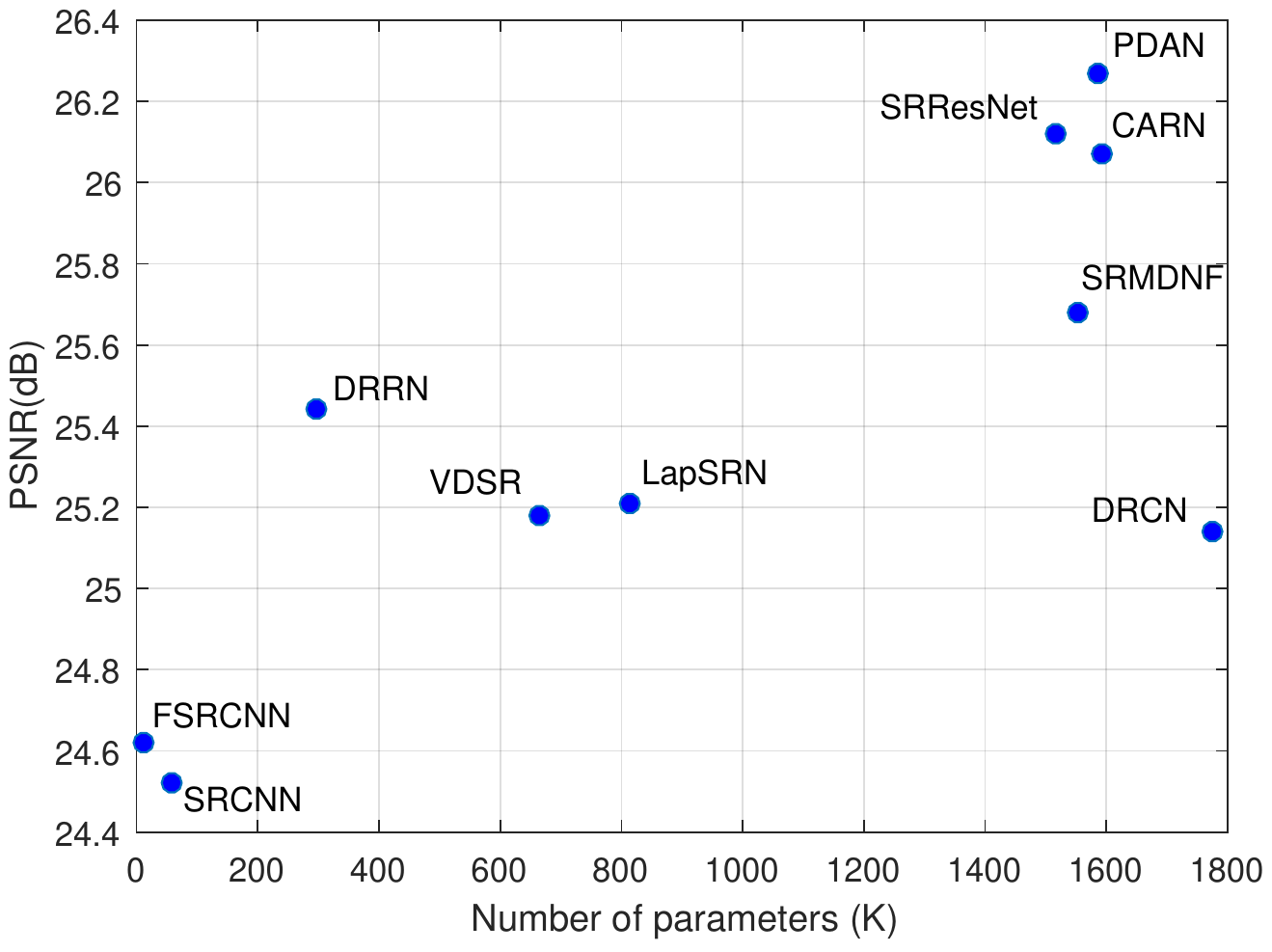}\\
	\caption{PSNR v.s. Parameters. The PSNR values are evaluated on Urban100 with scaling factor $\times 4$}\label{fig:sandian}
\end{figure}

\begin{table*}
	\newcommand{\tabincell}[2]{\begin{tabular}{@{}#1@{}}#2\end{tabular}}
	\centering
	\caption{Quantitative results with BD and DN degradation modules for scaling factor $ \times 3$. Red/blue text: best/second-best among all methods.}\label{comp_with_others_BD}
	\begin{tabular}{L{1.8cm}|C{0.8cm}|C{0.8cm}C{0.8cm}|C{0.8cm}C{0.8cm}|C{0.8cm}C{0.8cm}|C{0.8cm}C{0.8cm}|C{0.8cm}C{0.8cm}}
		\toprule
		\multicolumn{1}{c}{} & \multicolumn{1}{c}{} &
		\multicolumn{2}{c}{Set5} & \multicolumn{2}{c}{Set14} & \multicolumn{2}{c}{Bsd100} & \multicolumn{2}{c}{Urban100} & \multicolumn{2}{c}{Manga109}\\
		Methods & Model & PSNR & SSIM & PSNR & SSIM & PSNR & SSIM & PSNR & SSIM & PSNR & SSIM \\
		\midrule
		\midrule
		\multirow{2}{*}{Bicubic} & BD & 28.78 & 0.8308 & 26.38 & 0.7271 & 26.33 & 0.6918 & 23.52 & 0.6862 & 25.46 & 0.8149 \\
		& DN & 24.01 & 0.5369 & 22.87 & 0.4724 & 22.92 & 0.4449 & 21.63 & 0.4687 & 23.01 & 0.5381 \\
		\midrule
		\multirow{2}{*}{SRCNN \cite{dong}} & BD & 32.05 & 0.8944 & 28.80 & 0.8074 & 28.13 & 0.7736 & 25.70 & 0.7770 & 29.47 & 0.8924 \\
		& DN & 25.01 & 0.6950 & 23.78 & 0.5898 & 23.76 & 0.5538 & 21.90 & 0.5737 & 23.75 & 0.7148 \\
		\midrule
		\multirow{2}{*}{FSRCNN \cite{dong2016accelerating}} & BD & 26.23 & 0.8124 & 24.44 & 0.7106 & 24.86 & 0.6832 & 22.04 & 0.6745 & 23.04 & 0.7927 \\
		& DN & 24.18 & 0.6932 & 23.02 & 0.5856 & 23.41 & 0.5556 & 21.15 & 0.5682 & 22.39 & 0.7111 \\
		\midrule
		\multirow{2}{*}{VDSR \cite{kim2016accurate}} & BD & 33.25 & 0.9150 & 29.46 & 0.8244 & 28.57 & 0.7893 & 26.61 & 0.8136 & 31.06 & 0.9234 \\
		& DN & 25.20 & 0.7183 & 24.00 & 0.6112 & 24.00 & 0.5749 & 22.22 & 0.6096 & 24.20 & 0.7525 \\
		\midrule
		\multirow{2}{*}{IRCNN\_G \cite{zhang2017learning}} & BD & 33.38 & 0.9182 & 29.63 & 0.8281 & 28.65 & 0.7922 & 26.77 & 0.8154 & 31.15 & 0.9245 \\
		& DN & 25.70 & 0.7379 & 24.45 & 0.6305 & 24.28 & 0.5900 & 22.90 & 0.6429 & 24.88 & 0.7765 \\
		\midrule
		\multirow{2}{*}{IRCNN\_C \cite{zhang2017learning}} & BD & 33.17 & 0.9157 & 29.55 & 0.8271 & 28.49 & 0.7886 & 26.47 & 0.8081 & 31.13 & 0.9236 \\
		& DN & 27.48 & 0.7925 & 25.92 & 0.6932 & 25.55 & 0.6481 & 23.93 & 0.6950 & 26.07 & 0.8253 \\
		\midrule
		\multirow{2}{*}{SRMDNF \cite{zhang2018learning}} & BD & 34.09 & 0.9242 & 30.11 & 0.8364 & 28.98 & 0.8009 & 27.50 & 0.8370 & 32.97 & 0.9391 \\
		& DN & 27.74 & 0.8026 & 26.13 & 0.6974 & 25.64 & 0.6495 & 24.28 & 0.7092 & 26.72 & 0.8424 \\
		\midrule
		\multirow{2}{*}{PDAN (ours)} & BD & \textcolor{blue}{34.47} & \textcolor{blue}{0.9270} & \textcolor{blue}{30.43} & \textcolor{blue}{0.8425} & \textcolor{blue}{29.16} & \textcolor{blue}{0.8058} & \textcolor{blue}{28.28} & \textcolor{blue}{0.8533} & \textcolor{blue}{33.85} & \textcolor{blue}{0.9448} \\
		& DN & \textcolor{red}{28.51} & \textcolor{red}{0.8156} & \textcolor{blue}{26.57} & \textcolor{red}{0.7101} & \textcolor{red}{25.93} & \textcolor{red}{0.6600} & \textcolor{blue}{24.84} & \textcolor{blue}{0.7340} & \textcolor{blue}{27.88} & \textcolor{blue}{0.8574} \\
		\midrule
		\midrule
		\multirow{2}{*}{RDN \cite{zhang2018residual}} & BD & \textcolor{red}{34.58} & \textcolor{red}{0.9280} & \textcolor{red}{30.53} & \textcolor{red}{0.8447} & \textcolor{red}{29.23} & \textcolor{red}{0.8079} & \textcolor{red}{28.46} & \textcolor{red}{0.8582} & \textcolor{red}{33.97} & \textcolor{red}{0.9465} \\
		& DN & \textcolor{blue}{28.47} & \textcolor{blue}{0.8151} & \textcolor{red}{26.60} & \textcolor{blue}{0.7101} & \textcolor{blue}{25.93} & \textcolor{blue}{0.6573} & \textcolor{red}{24.92} & \textcolor{red}{0.7364} & \textcolor{red}{28.00} & \textcolor{red}{0.8591} \\
		\bottomrule
	\end{tabular}
\end{table*}

\subsection{Results with BD and DN Degradations}

Following \cite{zhang2018residual, zhang2017learning}, we also carry out experiments with blur-downscale (BD) and downscale-noise (DN) degradation models. The proposed method PDAN is compared with some state-of-the-art SR methods: SRCNN \cite{dong}, FSRCNN \cite{dong2016accelerating}, VDSR \cite{kim2016accurate}, IRCNN\_G \cite{zhang2017learning}, IRCNN\_C  \cite{zhang2017learning}, SRMDNF \cite{zhang2018learning}, and RDN \cite{zhang2018residual}. As shown in Table \ref{comp_with_others_BD}, our PDAN achieve the best performance on almost all quantitative results compared with other SR methods with scaling factor $ \times {\rm{3}}$. In particular, in comparison with RDN \cite{zhang2018residual}, our proposed PDAN still obtains comparable results. It is worth noting that PDAN has fewer parameters than RDN (1.6 M v.s. 22.3 M, 14$ \times $ less).

\subsection{Model size}

The number of parameters is a key factor for constructing a lightweight image SR model. From Table \ref{comp_with_others_BI}, we can observe that our method achieve better results than other stater-of-the-art lightweight methods. As shown in Fig. \ref{fig:sandian}, we illustrate the comparisons about parameters and PSNR on Urban100 with upscaling factor $\times 4$. In comparison with other methods, our PDAN obtains superior performance with comparable model size. It demonstrates that our method achieves a better trade-off between the performance and model size.

\section{Conclusion} \label{sec:Conclusion}

In this paper, we propose a lightweight pyramidal dense attention network (PDAN) for image SR. Specifically, in the proposed pyramidal dense attention block, the pyramidal dense connections can efficiently exploit feature information in different layers. Meanwhile, the adaptive group convolution whose group size increases linearly with layers is presented to relieve the parameter explosion. In addition, we introduce a new joint attention mechanism to recalibrate the feature responses more accurately by considering cross-dimension interaction in an efficient way. Extensive experiments on image SR demonstrate the superior performance of our PDAN in terms of quantitative and qualitative results.

\appendices

\ifCLASSOPTIONcaptionsoff
\newpage
\fi

\bibliographystyle{IEEEtran}   %>>>> makes bibtex use spiejour.bst
\bibliography{egbib}   %>>>> bibliography data in report.bib

\begin{thebibliography}{10}
\providecommand{\url}[1]{#1}
\csname url@samestyle\endcsname
\providecommand{\newblock}{\relax}
\providecommand{\bibinfo}[2]{#2}
\providecommand{\BIBentrySTDinterwordspacing}{\spaceskip=0pt\relax}
\providecommand{\BIBentryALTinterwordstretchfactor}{4}
\providecommand{\BIBentryALTinterwordspacing}{\spaceskip=\fontdimen2\font plus
\BIBentryALTinterwordstretchfactor\fontdimen3\font minus
  \fontdimen4\font\relax}
\providecommand{\BIBforeignlanguage}[2]{{%
\expandafter\ifx\csname l@#1\endcsname\relax
\typeout{** WARNING: IEEEtran.bst: No hyphenation pattern has been}%
\typeout{** loaded for the language `#1'. Using the pattern for}%
\typeout{** the default language instead.}%
\else
\language=\csname l@#1\endcsname
\fi
#2}}
\providecommand{\BIBdecl}{\relax}
\BIBdecl

\bibitem{zhou-et-al:scheme}
F.~Zhou, W.~Yang, and Q.~Liao, ``Interpolation-based image super-resolution
  using multisurface fitting,'' \emph{IEEE Transactions on Image Processing},
  vol.~21, no.~7, pp. 3312--3318, 2012.

\bibitem{tai-et-al:scheme}
Y.~W. Tai, S.~Liu, M.~S. Brown, and S.~Lin, ``Super resolution using edge prior
  and single image detail synthesis,'' in \emph{Proceedings of the IEEE
  Conference on Computer Vision and Pattern Recognition}, 2010.

\bibitem{timofte2014a+}
R.~Timofte, V.~De~Smet, and L.~Van~Gool, ``A+: Adjusted anchored neighborhood
  regression for fast super-resolution,'' in \emph{Asian Conference on Computer
  Cision}.\hskip 1em plus 0.5em minus 0.4em\relax Springer, 2014, pp. 111--126.

\bibitem{dong}
C.~Dong, C.~C. Loy, K.~He, and X.~Tang, ``Image super-resolution using deep
  convolutional networks,'' \emph{IEEE Transactions on Pattern Analysis and
  Machine Intelligence}, vol.~38, no.~2, pp. 295--307, 2015.

\bibitem{2016Deeply}
J.~Kim, J.~Kwon~Lee, and K.~Mu~Lee, ``Deeply-recursive convolutional network
  for image super-resolution,'' in \emph{Proceedings of the IEEE Conference on
  Computer Vision and Pattern Recognition}, 2016, pp. 1637--1645.

\bibitem{2017Image}
Y.~Tai, J.~Yang, and X.~Liu, ``Image super-resolution via deep recursive
  residual network,'' in \emph{Proceedings of the IEEE Conference on Computer
  Vision and Pattern Recognition}, 2017.

\bibitem{2017Deep}
Y.~Tang, W.~Gong, X.~Chen, and W.~Li, ``Deep inception-residual laplacian
  pyramid networks for accurate single image super-resolution,'' \emph{IEEE
  Transactions on Neural Networks and Learning Systems}, vol.~31, no.~5, 2017.

\bibitem{2017Enhanced}
B.~Lim, S.~Son, H.~Kim, S.~Nah, and K.~Mu~Lee, ``Enhanced deep residual
  networks for single image super-resolution,'' in \emph{Proceedings of the
  IEEE Conference on Computer Vision and Pattern Recognition Workshops}, 2017,
  pp. 136--144.

\bibitem{2020Wavelet}
J.~Xin, J.~Li, X.~Jiang, N.~Wang, H.~Huang, and X.~Gao, ``Wavelet-based dual
  recursive network for image super-resolution,'' \emph{IEEE Transactions on
  Neural Networks and Learning Systems}, vol.~PP, no.~99, pp. 1--14, 2020.

\bibitem{2020Multi}
H.~Wu, Z.~Zou, J.~Gui, W.~Zeng, J.~Ye, J.~Zhang, H.~Liu, and Z.~Wei,
  ``Multi-grained attention networks for single image super-resolution,''
  \emph{IEEE Transactions on Circuits and Systems for Video Technology},
  vol.~PP, no.~99, pp. 1--1, 2020.

\bibitem{2020Deep}
Z.~Wang, J.~Chen, and S.~C.~H. Hoi, ``Deep learning for image super-resolution:
  A survey,'' \emph{IEEE Transactions on Pattern Analysis and Machine
  Intelligence}, vol.~PP, no.~99, pp. 1--1, 2020.

\bibitem{he2016deep}
K.~He, X.~Zhang, S.~Ren, and J.~Sun, ``Deep residual learning for image
  recognition,'' in \emph{Proceedings of the IEEE Conference on Computer Vision
  and Pattern Recognition}, 2016, pp. 770--778.

\bibitem{huang2017densely}
G.~Huang, Z.~Liu, L.~Van Der~Maaten, and K.~Q. Weinberger, ``Densely connected
  convolutional networks,'' in \emph{Proceedings of the IEEE Conference on
  Computer Vision and Pattern Recognition}, 2017, pp. 4700--4708.

\bibitem{zhang2018residual}
Y.~Zhang, Y.~Tian, Y.~Kong, B.~Zhong, and Y.~Fu, ``Residual dense network for
  image super-resolution,'' in \emph{Proceedings of the IEEE Conference on
  Computer Vision and Pattern Recognition}, 2018, pp. 2472--2481.

\bibitem{dai2019second-order}
T.~Dai, J.~Cai, Y.~Zhang, S.~Xia, and L.~Zhang, ``Second-order attention
  network for single image super-resolution,'' in \emph{Proceedings of the IEEE
  Conference on Computer Vision and Pattern Recognition}, 2019, pp.
  11\,065--11\,074.

\bibitem{dong2016accelerating}
C.~Dong, C.~C. Loy, and X.~Tang, ``Accelerating the super-resolution
  convolutional neural network,'' in \emph{Proceedings of the European
  Conference on Computer Vision}.\hskip 1em plus 0.5em minus 0.4em\relax
  Springer, 2016, pp. 391--407.

\bibitem{shi2016real}
W.~Shi, J.~Caballero, F.~Husz{\'a}r, J.~Totz, A.~P. Aitken, R.~Bishop,
  D.~Rueckert, and Z.~Wang, ``Real-time single image and video super-resolution
  using an efficient sub-pixel convolutional neural network,'' in
  \emph{Proceedings of the IEEE Conference on Computer Vision and Pattern
  Recognition}, 2016, pp. 1874--1883.

\bibitem{zhang2018image}
Y.~Zhang, K.~Li, K.~Li, L.~Wang, B.~Zhong, and Y.~Fu, ``Image super-resolution
  using very deep residual channel attention networks,'' in \emph{Proceedings
  of the European Conference on Computer Vision}, 2018, pp. 286--301.

\bibitem{2018Fast}
N.~Ahn, B.~Kang, and K.~A. Sohn, ``Fast, accurate, and lightweight
  super-resolution with cascading residual network,'' in \emph{Proceedings of
  the European Conference on Computer Vision}, 2018.

\bibitem{2017Interleaved}
T.~Zhang, G.~J. Qi, B.~Xiao, and J.~Wang, ``Interleaved group convolutions for
  deep neural networks,'' 2017.

\bibitem{hui}
Z.~Hui, X.~Wang, and X.~Gao, ``Fast and accurate single image super-resolution
  via information distillation network,'' in \emph{Proceedings of the IEEE
  Conference on Computer Vision and Pattern Recognition}, 2018.

\bibitem{2019Lightweight}
Z.~Hui, X.~Gao, Y.~Yang, and X.~Wang, ``Lightweight image super-resolution with
  information multi-distillation network,'' in \emph{Acm International
  Conference}, 2019.

\bibitem{AIM9022144}
K.~{Zhang}, S.~{Gu}, R.~{Timofte}, Z.~{Hui}, X.~{Wang}, X.~{Gao}, D.~{Xiong},
  S.~{Liu}, R.~{Gang}, N.~{Nan}, C.~{Li}, X.~{Zou}, N.~{Kang}, Z.~{Wang},
  H.~{Xu}, C.~{Wang}, Z.~{Li}, L.~{Wang}, J.~{Shi}, W.~{Sun}, Z.~{Lang},
  J.~{Nie}, W.~{Wei}, L.~{Zhang}, Y.~{Niu}, P.~{Zhuo}, X.~{Kong}, L.~{Sun}, and
  W.~{Wang}, ``Aim 2019 challenge on constrained super-resolution: Methods and
  results,'' in \emph{2019 IEEE/CVF International Conference on Computer Vision
  Workshops}, 2019, pp. 3565--3574.

\bibitem{2017MobileNets}
A.~G. Howard, M.~Zhu, B.~Chen, D.~Kalenichenko, W.~Wang, T.~Weyand,
  M.~Andreetto, and H.~Adam, ``Mobilenets: Efficient convolutional neural
  networks for mobile vision applications,'' 2017.

\bibitem{2020Improving}
J.~J. Liu, Q.~Hou, M.~M. Cheng, C.~Wang, and J.~Feng, ``Improving convolutional
  networks with self-calibrated convolutions,'' in \emph{2020 IEEE/CVF
  Conference on Computer Vision and Pattern Recognition (CVPR)}, 2020.

\bibitem{2019Fast}
X.~Chu, B.~Zhang, H.~Ma, R.~Xu, J.~Li, and Q.~Li, ``Fast, accurate and
  lightweight super-resolution with neural architecture search,'' 2019.

\bibitem{hu2019channel}
Y.~Hu, J.~Li, Y.~Huang, and X.~Gao, ``Channel-wise and spatial feature
  modulation network for single image super-resolution,'' \emph{IEEE
  Transactions on Circuits and Systems for Video Technology}, 2019.

\bibitem{hu2018squeeze}
J.~Hu, L.~Shen, and G.~Sun, ``Squeeze-and-excitation networks,'' in
  \emph{Proceedings of the IEEE Conference on Computer Vision and Pattern
  Recognition}, 2018, pp. 7132--7141.

\bibitem{2018CBAM}
S.~Woo, J.~Park, J.~Y. Lee, and I.~S. Kweon, ``Cbam: Convolutional block
  attention module,'' in \emph{Proceedings of the European Conference on
  Computer Vision}, 2018.

\bibitem{2020MC}
Z.~Zhu, Z.~P. Bian, J.~Hou, Y.~Wang, and L.~P. Chau, ``When residual learning
  meets dense aggregation: Rethinking the aggregation of deep neural
  networks,'' 2020.

\bibitem{Misra_2021_WACV}
D.~Misra, T.~Nalamada, A.~U. Arasanipalai, and Q.~Hou, ``Rotate to attend:
  Convolutional triplet attention module,'' in \emph{Proceedings of the
  IEEE/CVF Winter Conference on Applications of Computer Vision (WACV)},
  January 2021, pp. 3139--3148.

\bibitem{2011Deep}
X.~Glorot, A.~Bordes, and Y.~Bengio, ``Deep sparse rectifier neural networks,''
  in \emph{Proceedings of the 14th International Conference on Artificial
  Intelligence and Statistics (AISTATS)}, 2011, pp. 315--323.

\bibitem{2016Multi}
F.~Yu and V.~Koltun, ``Multi-scale context aggregation by dilated
  convolutions,'' 2016.

\bibitem{2015Batch}
S.~Ioffe and C.~Szegedy, ``Batch normalization: Accelerating deep network
  training by reducing internal covariate shift,'' 2015.

\bibitem{timofte2017ntire}
R.~Timofte, E.~Agustsson, L.~Van~Gool, M.-H. Yang, and L.~Zhang, ``Ntire 2017
  challenge on single image super-resolution: Methods and results,'' in
  \emph{Proceedings of the IEEE Conference on Computer Vision and Pattern
  Recognition Workshops}, 2017, pp. 114--125.

\bibitem{bevilacqua2012low}
M.~Bevilacqua, A.~Roumy, C.~Guillemot, and M.~L. Alberi-Morel, ``Low-complexity
  single-image super-resolution based on nonnegative neighbor embedding,'' in
  \emph{Proceedings of the British Machine Vision Conference}, 2012.

\bibitem{2010On}
R.~Zeyde, M.~Elad, and M.~Protter, ``On single image scale-up using
  sparse-representations,'' in \emph{International Conference on Curves and
  Surfaces}.\hskip 1em plus 0.5em minus 0.4em\relax Springer, 2010, pp.
  711--730.

\bibitem{2002A}
D.~Martin, C.~Fowlkes, D.~Tal, and J.~Malik, ``A database of human segmented
  natural images and its application to evaluating segmentation algorithms and
  measuring ecological statistics,'' in \emph{IEEE International Conference on
  Computer Vision}, 2002.

\bibitem{2015Single}
J.-B. Huang, A.~Singh, and N.~Ahuja, ``Single image super-resolution from
  transformed self-exemplars,'' in \emph{Proceedings of the IEEE Conference on
  Computer Vision and Pattern Recognition}, 2015, pp. 5197--5206.

\bibitem{2017Sketch}
Y.~Matsui, K.~Ito, Y.~Aramaki, A.~Fujimoto, T.~Ogawa, T.~Yamasaki, and
  K.~Aizawa, ``Sketch-based manga retrieval using manga109 dataset,''
  \emph{Multimedia Tools and Applications}, vol.~76, no.~20, pp.
  21\,811--21\,838, 2017.

\bibitem{wang2004image}
Z.~Wang, A.~C. Bovik, H.~R. Sheikh, E.~P. Simoncelli \emph{et~al.}, ``Image
  quality assessment: from error visibility to structural similarity,''
  \emph{IEEE Transactions on Image Processing}, vol.~13, no.~4, pp. 600--612,
  2004.

\bibitem{2014Adam}
D.~Kingma and J.~Ba, ``Adam: A method for stochastic optimization,''
  \emph{Computer Science}, 2014.

\bibitem{paszke2017automatic}
A.~Paszke, S.~Gross, S.~Chintala, G.~Chanan, E.~Yang, Z.~DeVito, Z.~Lin,
  A.~Desmaison, L.~Antiga, and A.~Lerer, ``Automatic differentiation in
  pytorch,'' 2017.

\bibitem{kim2016accurate}
J.~Kim, J.~Kwon~Lee, and K.~Mu~Lee, ``Accurate image super-resolution using
  very deep convolutional networks,'' in \emph{Proceedings of the IEEE
  Conference on Computer Vision and Pattern Recognition}, 2016, pp. 1646--1654.

\bibitem{lai2017deep}
W.-S. Lai, J.-B. Huang, N.~Ahuja, and M.-H. Yang, ``Deep laplacian pyramid
  networks for fast and accurate super-resolution,'' in \emph{Proceedings of
  the IEEE Conference on Computer Vision and Pattern Recognition}, 2017, pp.
  624--632.

\bibitem{zhang2018learning}
K.~Zhang, W.~Zuo, and L.~Zhang, ``Learning a single convolutional
  super-resolution network for multiple degradations,'' in \emph{Proceedings of
  the IEEE Conference on Computer Vision and Pattern Recognition}, 2018, pp.
  3262--3271.

\bibitem{ledig2017photo}
C.~Ledig, L.~Theis, F.~Husz{\'a}r, J.~Caballero, A.~Cunningham, A.~Acosta,
  A.~Aitken, A.~Tejani, J.~Totz, Z.~Wang \emph{et~al.}, ``Photo-realistic
  single image super-resolution using a generative adversarial network,'' in
  \emph{Proceedings of the IEEE Conference on Computer Vision and Pattern
  Recognition}, 2017, pp. 4681--4690.

\bibitem{zhang2017learning}
K.~Zhang, W.~Zuo, S.~Gu, and L.~Zhang, ``Learning deep cnn denoiser prior for
  image restoration,'' in \emph{Proceedings of the IEEE Conference on Computer
  Vision and Pattern Recognition}, 2017, pp. 3929--3938.

\end{thebibliography}

\end{document}